\documentclass{aa}  
\usepackage{psfig,graphicx,epsfig}
\usepackage{lscape}
\usepackage{natbib}
\bibpunct{(}{)}{;}{a}{}{,}
\newcommand\PSR{PSR J1740--5340}
\begin{document}
   \title{X-rays from the eclipsing millisecond pulsar PSR J$1740-5340$ in 
   the globular cluster NGC 6397}
   \titlerunning{X-rays from the eclipsing millisecond pulsar PSR J1740--5340}
   \authorrunning{Huang \& Becker}
   \author{R. H. H. Huang \& W. Becker}
   \institute{Max-Planck-Institut f\"ur extraterrestrische Physik,
               Giessenbachstrasse 1, 85748 Garching, Germany}

\abstract 
{}
{The millisecond pulsar PSR J$1740-5340$ in the globular cluster NGC\,6397 
shows radio eclipses over $\sim40\%$ of its binary orbit. A first Chandra observation revealed 
indications for the X-ray flux being orbit dependent as well. In this work, we analysed five 
datasets of archival Chandra data taken between 2000 and 2007 to investigate the 
emission across the pulsar's binary orbit.} 
{Utilizing archival Chandra observations of PSR J1740--5340, we performed a systematic 
timing and spectral analysis of this binary system.} 
{Using a $\chi^{2}$-test, the significance for intra-binary orbital modulation was found to be between 
88.5\% and 99.6\%, depending on the number of phase bins used to construct the light curve. 
Applying the unbiased statistical Kolmogorov-Smirnov (KS) test did not indicate any significant 
intra-binary orbital modulation. However, comparing the counting rates observed
at different epochs, a flux variability on times scales of days to years is indicated. 
The possible origin of the X-ray emission is discussed in a number of different scenarios.} 
{}

\maketitle


\section{Introduction}

The millisecond pulsar (MSP) PSR J1740--5340 (6397A), located in the core-collapsed 
globular cluster NGC\,6397 at a distance of 2.5 kpc, was discovered during a systematic search for 
ms-pulsars in Galactic globular clusters using the Parkes Radio Telescope 
\citep{damico2001}. It is orbiting around a massive late type companion 
\citep[$>$ 0.14 $M_{\odot}$;][]{2003orosz} with an orbital period of 1.35 day 
\citep{damico2001b}. The inclination of the binary system to the line of sight is 
$i\sim 50^\circ$ \citep{2003orosz}. The pulsar's spin period and period derivative 
are $P=3.65$ ms \citep{damico2001} and $\dot{P} = 4.0 \times 10^{-20} ~\mathrm{s ~s^{-1}}$ 
\citep{2005possenti}. The contribution to $\dot{P}$  due to possible pulsar 
acceleration in the cluster's gravitational field is estimated to be smaller 
than $10^{-20}~\mathrm{s ~s^{-1}} $  by its large offset of 0.6 arcmin (corresponding
to eleven core radii) from the cluster center \citep{damico2001b}. The spin 
parameters imply a spin-down luminosity, a characteristic age and a dipole 
surface magnetic field of \PSR\ of $\dot{E} \sim 3.3 \times 10^{34} 
~\mathrm{ergs ~s^{-1}}$, $\tau_{c} = P/2\dot{P} \sim 1.4 \times 10^{9}$ years, 
and $B_\perp \sim 3.9 \times 10^{8}$ G, respectively \citep{2005possenti}.  

The radio emission from \PSR\ is partially eclipsed in the orbital phase interval $0.05-0.45$ 
for approximately 40\% of each orbit \citep{damico2001,damico2001b}. Moreover, strong 
fluctuations of the radio signals were observed at nearly all orbital phases 
\citep{damico2001b}, which led to the interpretation that the pulsar is orbiting 
within an extended envelope of matter released from the companion. The optical light 
curve of the companion showing tidal distortions provides strong evidence that 
\PSR\ is orbiting a companion whose Roche lobe is nearly completely filled \citep{2001ferraro}. 
However, the system is in a radio-ejection phase in which accretion is inhibited by 
the radiation pressure exerted by the pulsar on the infalling matter. The strong 
interaction between the MSP flux and the plasma wind would explain the irregularities 
seen in the radio signals from \PSR\, \citep{2001ferraro,2002burderi}.

\PSR\ is in a wide orbit with a separation of $\sim 6.5\,R_{\odot}$, so that the wind 
energy density impinging on the non-degenerate companion should be significantly less 
than that estimated for other much tighter eclipsing binary systems, such as the field 
ms-pulsar PSR B1957+20 which has a separation of $\sim 0.04~R_{\odot}$ 
from its companion \citep{1994arzoumanian}. 
\PSR\ therefore is unlikely to drive a wind of sufficient density off its companion 
\citep{damico2001b}. \citet{2003orosz} have studied the optical light curve of the companion 
and found no evidence of heating from the pulsar radiation, which supports the aforementioned 
inference. \citet{2003sabbi} have investigated the chemical composition of the non-degenerate 
companion star and found a strong depletion of carbon. This suggests a scenario 
in which the companion is an evolved star that has lost most of its surface layers. In view of the 
high rotational velocity of the companion \citep[$\sim50$~km/s;][]{2003sabbi}, the stellar 
wind possibly can be strong enough to cause the mass-loss and result in an extended envelope 
of matter. 

X-ray emission from this binary system was detected with the Chandra X-ray Observatory 
\citep{2001grindlay}. Based on an observation in 2000 (cf. Table 1) which only covered 
$\sim 40\%$ of the binary orbit, \citet{2002grindlay} reported evidence that the count 
rate of the system appears to increase by a factor of $\sim2$ at phase 0.4, just before 
the pulsar comes out of the radio eclipse. However, the short exposure time and the limited 
orbital phase coverage of this first observation (cf. top panel in Figure~1 and Table~1) 
did not support a detailed temporal and spectral analysis of the emission.

In this publication we report on a search for X-ray orbital modulation as well as on 
a spectral analysis of the PSR J1740--5340 binary system making use of archival 
Chandra data that cover various orbital phases ranges. 


\section{Observations and data analysis}

\begin{figure}
   \centering
   \includegraphics[angle=0,width=9.3cm]{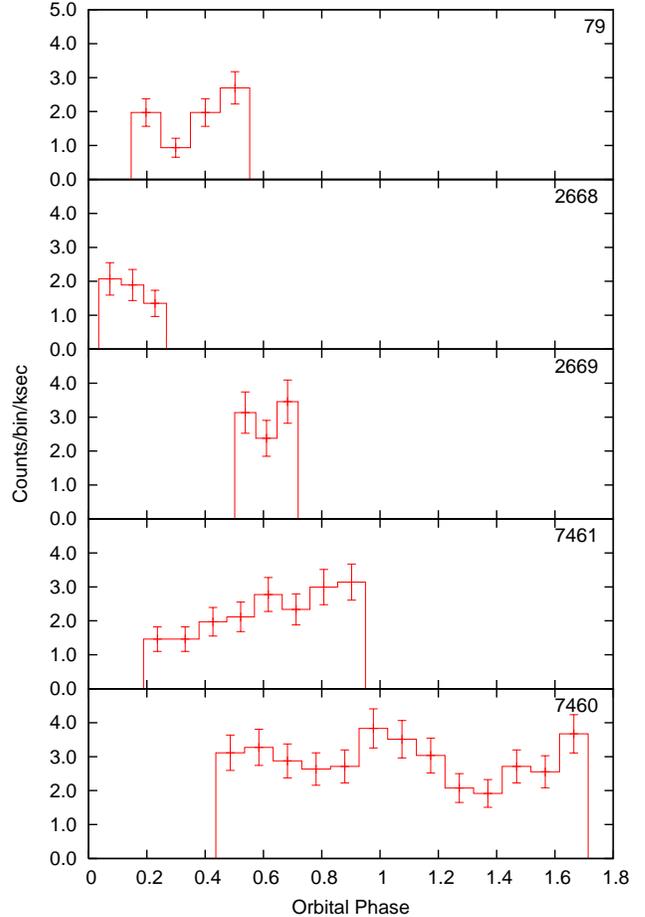}
      \caption{Orbital phase coverages of five Chandra observations targeted on the binary system \PSR. 
      The Y-axis shows the X-ray source counts per bin per kilosecond in the energy range 
      $0.3-8.0$ keV. The X-axis is the binary orbital phase \citep[from ephemeris of][]{damico2001b}.
      $\phi$=0.0 is at the ascending node of the pulsar orbit.
      Binning is so that the signal-to-noise ratio in each bin is $\ge 3$.
      Error bars represent the $1\sigma$ uncertainties. All X-ray light curves are corrected 
      for vignetting and the various instrumental sensitivities of the different observing modes.
      Labels are the Obs. IDs (cf. Table~1).} 
   \label{allorb}
   \end{figure}

In total, five observations were targeted on the \PSR\ binary system using 
the Chandra Advanced CCD Imaging Spectrometer (ACIS). The first observation was 
performed on 2000 July 31 using the front-illuminated (FI) ACIS-I3 chip, while 
the other four observations taken on 2002 May 13 and 15, 2007 June 26, and July 
16 used the back illuminated (BI) chip ACIS-S3. We summarize the basic information 
of these observations in Table 1. Their binary orbit coverages are shown in 
Figure~\ref{allorb}. As can be seen from this figure, only Obs. IDs 7460 and 7461 
cover a larger fraction of the binary orbit. 
In the search for intra-orbital flux modulation and spectral variation, we therefore considered only these two datasets 
which have higher photon statistics and longer exposure times, while the other archival data were used 
to search for a possible flux variability on time scales of days or years. 
Data analysis was restricted to the energy range $0.3-8.0$\,keV. Searching for X-ray pulses 
from \PSR\ was precluded by the inappropriate temporal resolution of the observing modes used.  
Standard processed level-2 data were used. Correction for aspect offset was applied before analysis. 
All data were processed with Chandra Interactive Analysis Observations (CIAO) version 3.4 software and CALDB 
version 3.4.2.

\begin{center}
\begin{table*}
\begin{minipage}[bt]{1.0\linewidth}
\caption{List of Chandra observations of PSR J1740--5340 in NGC 6397}
\label{obslog} \renewcommand{\footnoterule}{}  
\begin{tabular}{lccccccc}\hline \hline\\[-2ex]
      Obs. Date  &   Obs. ID &   Instrument &   Data Mode &   Effective Exp. Time &    Orbital Phase  &    Net count rate$^a$ & Off-axis(') \\\hline\\[-2ex]
     2000-07-31  &     79    &    ACIS-I    &       Faint &         48.3 ks       &   $0.146-0.553$   &    $1.89 \pm 0.25^b$  &  0.8745 \\[+0.5ex]                      
     2002-05-13  &   2668    &    ACIS-S    &       Faint &         28.1 ks       &   $0.035-0.267$   &    $1.74 \pm 0.25$    &  0.9368 \\[+0.5ex]
     2002-05-15  &   2669    &    ACIS-S    &       Faint &         26.7 ks       &   $0.501-0.719$   &    $2.89 \pm 0.33$    &  0.9214 \\[+0.5ex]
     2007-06-22  &   7461    &    ACIS-S    &  Very Faint &         90.0 ks       &   $0.189-0.950$   &    $2.26 \pm 0.16$    &  0.3732 \\[+0.5ex]
     2007-07-16  &   7460    &    ACIS-S    &  Very Faint &        149.6 ks       &   $0.437-1.714^c$ &    $2.92 \pm 0.14$    &  0.2129 \\[+0.5ex]\hline\\[-2ex]
\end{tabular}
\end{minipage}
${}^a$ACIS-S net count rates in units of $10^{-3}$ counts/sec corrected to the on-axis.\\
${}^b$The ACIS-I on-axis net count rate is 1.31 $\pm$ 0.17.  In order to derive a comparable 
      ACIS-S net count rate from the observed\\ ACIS-I rate we convert the ACIS-I rate by using 
      the WebPIMMS tool (http://heasarc.gsfc.nasa.gov/Tools/w3pimms.html).\\
${}^c$The total exposure time covers more than one binary orbit.
\end{table*}
\end{center}

\subsection{Search for orbital modulation}

In order to search for a modulation of the X-ray flux as a function of orbital phase, we 
extracted the photons from a circle of 2 arcsecond radius centered on the radio 
pulsar position RA (J2000) $= 17^{h}40^{m}44\fs59$, Dec $= -53^{\circ}40'40\farcs9$ 
\citep{damico2001b}. In the 2007 observations, $\sim90\%$ of all source counts are located 
within this cut radius. The photon arrival times were corrected to the solar system 
barycenter with the CIAO tool AXBARY (JPL DE200 solar system ephemeris). We used the 
pulsar ephemeris from \citet{damico2001b} and correct the photon arrival times for the 
orbital motion of the binary system \citep{1976blandford}. Since \PSR\ is far out of 
the globular cluster core, the correction for gravitational acceleration in the 
cluster potential was ignored. 

\begin{figure}
   \centering
   \includegraphics[angle=-90,width=9.3cm]{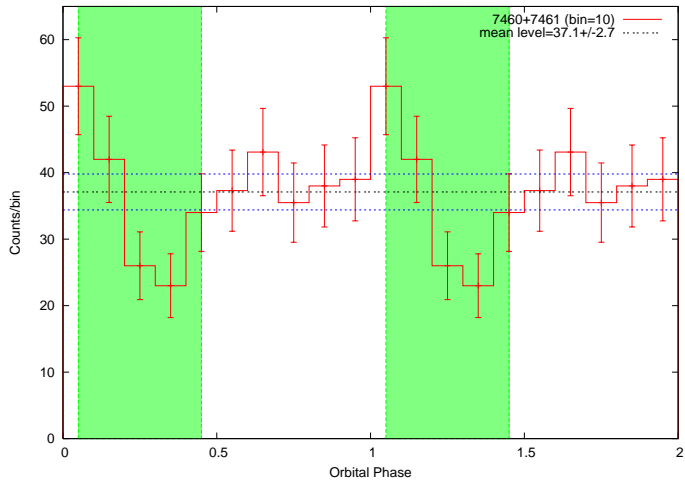}
      \caption{X-ray source counts (0.3--8.0 keV) vs. orbital binary phase according to the
      ephemeris of \citet{damico2001b}. Two orbital cycles are shown for clarity. The 
      background noise level is found to be at $\sim 0.35$ counts/bin.  $\phi = 0.0$ 
      corresponds to the ascending node of the pulsar orbit. Error bars indicate the 
      $1\sigma$ uncertainty. The shaded regions between the orbital phases 0.05--0.45
      and 1.05--1.45 mark the radio eclipse of the pulsar PSR~J1740--5340.
      The dotted lines indicates the best-fit constant level and its $1\sigma$ uncertainty.}
   \label{ltc}
   \end{figure}

We tested the 2007 data for intra-orbital flux modulation by fitting a constant to the X-ray light curve. 
In order to have the signal-to-noise ratio for each phase bin higher than $\sim$ 4, the numbers of 
phase bins were restricted to be within 10 to 15 bins per orbital period. 
In addition, we tested 20 and 40 bins per orbital period for comparison. 
Using a $\chi^{2}$-test, the significance for a flux modulation over the observed orbit was found to be 
between 88.5\% and 99.6\%, depending on the number of phase bins used to construct the light curve. 
Table~\ref{chi2} summarizes the results for various bin numbers. The X-ray light curve for 10 phase 
bins is shown in Figure \ref{ltc}.

\begin{table}
\caption{Significance for intra-orbital flux modulation by fitting the
 light curves of different phase bins against a constant by using 
 the $\chi^{2}$ test statistics. The best-fit mean levels are listed as well.}
\label{chi2}
\centering
\begin{tabular}{cccc}
    \hline \hline
      Number of bins & Mean level & $\chi^{2}_{\nu}$ & Significance(\%) \\\hline\\[-1.5ex]
      10 &  37.1$\pm$2.7 & 2.22 & 98.2 \\[+0.5ex] 
      11 &  33.4$\pm$2.4 & 1.82 & 94.8 \\[+0.5ex]
      12 &  29.6$\pm$2.1 & 1.62 & 91.2 \\[+0.5ex]
      13 &  28.6$\pm$2.1 & 2.38 & 99.6 \\[+0.5ex]
      14 &  26.4$\pm$2.0 & 1.86 & 97.0 \\[+0.5ex]
      15 &  24.7$\pm$1.9 & 2.01  & 98.7 \\[+0.5ex]
      20 &  18.4$\pm$1.1 & 1.34 & 89.0 \\[+0.5ex]
      40 &  9.2$\pm$0.5  & 1.25  & 88.5 \\[+0.5ex]
\hline
\end{tabular}
\end{table}

\begin{figure}
   \hspace{0.5cm}\includegraphics[angle=-90,width=10.5cm]{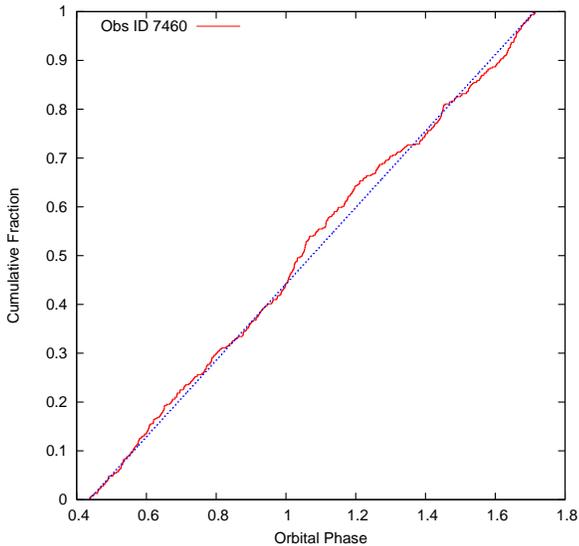}
      \caption{Kolmogorov-Smirnov test. The cumulative distribution function
       from the observed data is plotted in red and the blue dotted
       line shows the estimated cumulative distribution generated by assuming
       a constant X-ray flux. }
   \label{kstest}
   \end{figure}

The disadvantage of the $\chi^{2}$-test is its dependence on the number of phase bins. We
therefore applied the Kolmogorov-Smirnov (KS) test to the unbinned light curve data in 
order to have a bin-independent statistical evaluation of the X-ray emission variability 
(see Figure~\ref{kstest}). Calculating the corresponding KS probabilities between our data 
set and the cumulative distribution function generated by assuming a constant X-ray flux
indicated no significant deviation between these two distributions. The significance for 
an intra-orbital flux modulation from this test is only at the level of $\sim$68\%. A similar
result is obtained if we restrict the analysis to the hard energy band above 2 keV.

\subsection{Spectral analysis}

To investigate whether the X-ray spectral behavior of \PSR\ varies across 
the orbit, we analysed the X-ray spectrum within and outside the eclipsing region separately. 
The source and background spectra were extracted from the same region as mentioned in \S2.1 
and from a source-free region close to the pulsar. Response files for the corresponding 
observations were created by using the tools MKACISRMF and MKARF of CIAO. The background-subtracted 
count rates from the PSR J$1740-5340$ binary system per observation are given in Table~\ref{obslog}. 
Extracted spectra were dynamically binned in accordance with the photon statistic in each dataset.
Model spectra to Obs. IDs. 7460 and 7461 datasets were fitted using XSPEC 11.3.2. There are $183 \pm 14$ 
and $457 \pm 21$ net counts in total for the spectral analysis inside and outside the eclipsing 
region, respectively. 

Assuming that the emission originates from the shock interaction of the pulsar wind with 
the wind of the companion star or from non-thermal emission processes in the pulsar
magnetosphere, we expect the X-radiation to be synchrotron. To test this hypothesis,  
we fitted the spectrum with an absorbed power-law model (PL). For the PL fit, no significant 
variation of the spectral parameters is found between the spectrum inside and outside the 
eclipsing region. 
A thermal bremsstrahlung model (TB) was also tested, which physically implies that the X-ray emission 
is either from a hot plasma presented in the binary system or from the free-free radiation of the companion's corona. 
We also tested whether a single blackbody (BB) can provide an appropriate modeling of the data.
However, this BB model cannot provide any statistically acceptable description of the observed spectra 
(i.e. $\chi^{2}_{\nu} > 2$) either inside the eclipsing region or outside.  

The spectral parameters inferred from these fits are summarized in Table~\ref{spec}. 
The spectral parameters inferred from the PL/TB fits are found to be consistent with those reported 
by \citet{2001grindlay,2002grindlay}. 
The best-fitted $N_{H}$ for each spectrum varies marginally, but is consistent with the values of 
$\sim1.03$ and $\sim2.22 \times 10^{21} \mathrm{cm^{-2}}$ inferred from the optical 
reddening of NGC 6397 and from the radio dispersion measure (DM) of \PSR, respectively.

\begin{table*}
\caption{Parameters from spectral fits to the PSR J$1740-5340$ binary system.}
\label{spec}
\begin{center}
\renewcommand{\footnoterule}{}  
\begin{tabular}{lccccccc}
    \hline \hline\\[-1.5ex]
      Orbital Phase &
      Model $^{a}$ &
      $N_{H}$ $^{b}$ &
      $\Gamma$/kT(keV) &
      $\chi^{2}_{\nu}$/dof &
      $F_{x}$ $^{c}$ &
      $L_{x}$ $^{d}$ & 
      Note \\[0.5ex]
\hline\\[-1ex]
      0.05--0.45 & PL & $<1.4$ & $1.8^{+0.4}_{-0.3}$ & 1.39/20 & $1.9^{+1.8}_{-0.7}$ &  
                       $1.4^{+1.3}_{-0.5}\times10^{31}$ & inside eclipsing region \\
                    & TB & $<0.8$ & $7.2^{+10.4}_{-3.4}$ & 1.42/20 & $1.7^{+0.5}_{-0.4}$ & $1.3^{+0.4}_{-0.3}\times10^{31}$ & \\[0.5ex]
\hline\\[-1.5ex]
      0.45--0.05 & PL & $1.8^{+1.0}_{-0.7}$ & $1.7^{+0.1}_{-0.2}$ & 0.96/19 & $3.3^{+1.4}_{-0.8}$ & 
                       $2.5^{+1.1}_{-0.6}\times10^{31}$ & outside eclipsing region \\
                    & TB & $1.3^{+0.6}_{-0.9}$ & $7.3^{+9.6}_{-1.7}$ & 1.00/19 &  $2.9^{+0.4}_{-0.5}$ & 
                       $2.2^{+0.3}_{-0.4}\times10^{31}$ & \\[0.5ex]
\hline\\[-1.5ex]
      0.0--1.0 & PL & $0.6^{+0.4}_{-0.3}$ & $1.5^{+0.2}_{-0.1}$ & 1.07/19 & $2.8^{+0.9}_{-0.8}$ & 
                       $2.1^{+0.7}_{-0.6}\times10^{31}$  & whole orbit \\
                    & TB & $<0.7$ & $16.1^{+23.2}_{-7.6}$ & 1.11/19 &  $2.6^{+0.3}_{-0.4}$ & 
                       $1.9^{+0.2}_{-0.3}\times10^{31}$ & \\[0.5ex]
\hline \hline
\end{tabular}
\end{center} 
\begin{minipage}[t]{0.98\linewidth}
Note: Col. 1 gives the orbital phase, Cols. 2--5 the spectral model, the hydrogen
column density $N_{H}$, the best-fitted photon index or temperature, and the reduced $\chi^{2}$ together with degrees 
of freedom. Cols. 6--7 list the unabsorbed X-ray flux and luminosity in the $0.3-8.0$ keV energy range. Quoted errors 
indicate the 68\% confidence level for one parameter of interest. \\
\noindent 
a. PL $=$ powerlaw; TB $=$ thermal bremsstrahlung \\
b. In units of $10^{21}$~cm$^{-2}$ \\
c. Unabsorbed X-ray flux in units of $10^{-14}$~ergs~cm$^{-2}$~s$^{-1}$ and in the energy range of 0.3--8.0 keV. \\
d. Unabsorbed X-ray luminosity in units of ergs~s$^{-1}$ and in the energy range of 0.3--8.0 keV for a distance of 2.5 kpc  \citep{2001grindlay}. 
\end{minipage}
\end{table*}


\section{Summary and discussion}

\begin{figure}
   \includegraphics[angle=-90,width=9cm]{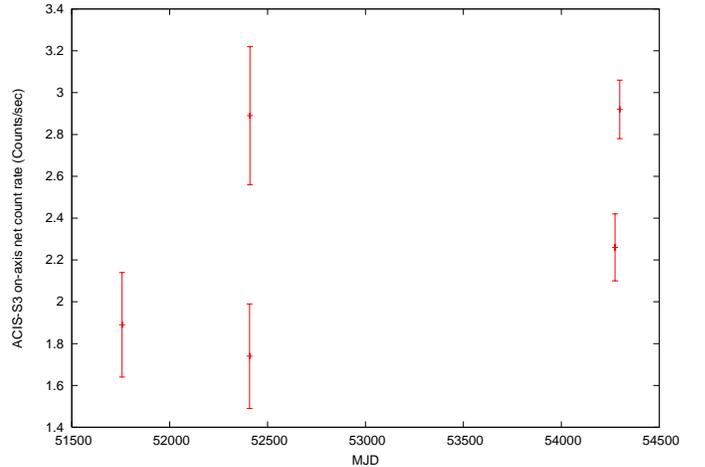}
      \caption{ACIS-S3 count rates of the \PSR\ binary system vs. observation dates.}
   \label{long}
   \end{figure}

We have searched for the orbital modulation of the X-ray emission from \PSR\ associated 
with NGC 6397. Only one of five archival datasets cover a full binary orbit. Analysing 
this data with a Kolmogorov-Smirnov test revealed no significant intra-orbital flux 
modulation. However, correlating the ACIS-S vignetting-corrected net counting rates 
observed at various orbital phases in 2000, 2002 and 2007 and assuming that the system 
in general shows no intra-orbital flux modulation reveals a $\sim 3\sigma$ flux variability 
on time scales of days to years (cf.~Col.~7 in Table 1). Figure~\ref{long} depicts the ACIS-S net 
count rates observed in the various observations. 

In the following, we will discuss three scenarios which we believe could be the main 
source of the observed X-rays in the \PSR\ binary system.

\subsection*{1. The Companion star}
The companion is observed to be a late-type star. It has an unusual position in the color-magnitude 
diagram (CMD) as being as luminous as a main-sequence turn-off star but redder \citep{2001ferraro,2003orosz}. 
This makes it quite similar to the ``red straggler'' active binaries which are
identified to be X-ray sources in other globular clusters \citep{2001albrow,2003edmonds1,2003edmonds2,2008bassa}. 
Detailed studies of X-ray sources in various globular clusters by \cite{2004bassa}
and \cite{2008verbunt} have shown that the separatrix 
$\log L_{\rm 0.5-2.5 keV}=34.0-0.4M_{\rm V}$ allows one to distinguish between the 
population of cataclysmic variables and that of magnetically active binaries. 
Here, $M_{\rm V}$ is the absolute V-band magnitude, which for the PSR J$1740-5340$ 
binary system was observed to be $M_{\rm V} \sim 5$ \citep{2003orosz}. Taking this 
together makes the properties observed from the companion star consistent with 
that from rapid rotating stars. The X-ray luminosities of late-type stars (i.e. F7 to M5) are found to be well
correlated with their equatorial rotational velocities $v_{\rm rot}$, according
to $L_{\mathrm{X}}\simeq10^{27}v^{2}_{\rm rot}$ ergs~s$^{-1}$ \citep{1981pallavicini}.
Adopting the $v_{\rm rot}\sim50$~km/s as 
reported by \citet{2003sabbi} yields an expected X-ray luminosity of $\sim2.5 \times10^{30}$~ergs~s$^{-1}$. 
Given that the $L_{\mathrm{X}}$ vs. $v^{2}_{\rm rot}$ correlation has a scatter of one order 
of magnitude \citep{1981pallavicini} it cannot be excluded that the X-ray emission
observed from the \PSR\ binary is coronal emission from the companion star only.
No significant X-ray intra-orbital modulation is observed, which is unlike what we have seen in the optical bands. 
The optical emission is expected to come from the surface of the companion star while the X-ray 
emission comes from the corona. As the companion star is interpreted to be tidal-distorted 
\citep{2003orosz}, the emitting area responsible for the optical intensity appears to be modulated. 
However, the X-ray emission is coronal and may be less or not affected by this distortion. 

\subsection*{2. The millisecond pulsar}
Considering a possible contribution from the pulsar magnetosphere is justified by the
pulsar's spin-down energy of $\dot{E} \sim 3.3 \times10^{34} ~\mathrm{ergs~s^{-1}}$. 
By assuming the best-fit X-ray conversion efficiency found for other field and millisecond 
pulsars, $L_\mathrm{X} \sim 10^{-3}-10^{-4} \dot{E}$ \citep{2009becker}, the spin-down power 
of \PSR\ implies an X-ray luminosity which is of the order of $\sim 10^{30} - 10^{31}~\mathrm{ergs ~s^{-1}}$,
in agreement with the observed luminosity (cf.~Table 3). Observing the pulsar in a mode 
suitable to resolve its X-ray pulses would be worthwhile and could constrain this interpretation.

\subsection*{3. Intrabinary shock emission}
\citet{2002grindlay} suggested that the X-ray emission they detected from the \PSR\ binary system
is due to the interaction between the pulsar and shocked gas lifted from the stellar companion.  
In the shock, the extended matter in the orbit is compressed and gives rise to a power law 
distribution of synchrotron-emitting particles, $N(\gamma)\propto\gamma^{-p}$, by accelerating a 
fraction of these particles to very high energies, where $\gamma$ is the Lorentz factor of 
the emitting particles. This should lead to an enhancement of X-ray emission from the system 
as the pulsar enters the eclipsing region and a subsequent decrease due to the increased 
absorption in this region. \citet{cheng2006} found that the X-ray luminosity 
of intrabinary shock emission in the 2--10~keV band can be described by $L^{\rm th}_{X}\sim
(0.3-1.2)\times10^{32}D_{11}^{-(p-1)/2}\dot{E}_{35}^{(p+5)/4}$~ergs~s$^{-1}$, assuming a Lorentz factor 
for the synchrotron-emitting particles of $\gamma=(0.5-1.0)\times10^{6}$, the fractional energy density of 
electrons to be $\epsilon_{e}=0.5$ and the fractional energy density of the magnetic field 
to be $\epsilon_{B}=0.02-0.1$. $p$ and $D_{11}$ denote the index of the postshock electron energy distribution 
$N(\gamma)\propto\gamma^{-p}$, and the orbital separation in units of $10^{11}$~cm. The X-ray 
photon index inferred from the spectrum is $\Gamma\sim1.8$ which would suggest that the emission from 
the shock is in a slow cooling regime \citep{cheng2006}. The photon index is related to 
$p$ as $\Gamma=(p+1)/2$. The observed $\Gamma$ implies that $p$ is $\sim2.6$. 
With $\dot{E}_{35}=0.33$ and $D_{11}=4.5$, $L^{\rm th}_{X}$ is found to be 
in the range of $\sim (0.1-4.4) \times10^{30}$~ergs~s$^{-1}$, which is at the observed level. 
Intrabinary shock emission thus can explain the observed X-ray luminosity though the
modulation depth and amplitude, which is also a function of the inclination angle, 
is unconstrained from this model.

Based on the preceding discussion, it is not possible to firmly distinguish the possible 
emission scenarios which may, alone or in combination, be responsible for the observed
X-ray emission. To disentangle them would require detailed orbital-phase resolved 
spectroscopy along with a search for X-ray pulses from the millisecond pulsar. Further 
dedicated X-ray observations covering more binary orbits are certainly needed to further
constrain the emission processes at work in this binary system.

\begin{acknowledgements}
This work made use of the Chandra data archive. The first author thanks C.Y. Hui and M. G\"{u}del 
for providing helpful suggestions and also acknowledges the receipt of funding provided 
by the Max-Planck Society in the frame work of the International Max-Planck Research School on 
Astrophysics at the University of Munich. 
\end{acknowledgements}

\bibliographystyle{aa}
\bibliography{bib6397}

\begin{thebibliography}{20}
\expandafter\ifx\csname natexlab\endcsname\relax\def\natexlab#1{#1}\fi

\bibitem[{{Albrow} {et~al.}(2001){Albrow}, {Gilliland}, {Brown}, {Edmonds},
  {Guhathakurta}, \& {Sarajedini}}]{2001albrow}
{Albrow}, M.~D., {Gilliland}, R.~L., {Brown}, T.~M., {et~al.} 2001, \apj, 559,
  1060

\bibitem[{{Arzoumanian} {et~al.}(1994){Arzoumanian}, {Fruchter}, \&
  {Taylor}}]{1994arzoumanian}
{Arzoumanian}, Z., {Fruchter}, A.~S., \& {Taylor}, J.~H. 1994, \apjl, 426, L85

\bibitem[{{Bassa} {et~al.}(2004){Bassa}, {Pooley}, {Homer}, {Verbunt},
  {Gaensler}, {Lewin}, {Anderson}, {Margon}, {Kaspi}, \& {van der
  Klis}}]{2004bassa}
{Bassa}, C., {Pooley}, D., {Homer}, L., {et~al.} 2004, \apj, 609, 755

\bibitem[{{Bassa} {et~al.}(2008){Bassa}, {Pooley}, {Verbunt}, {Homer},
  {Anderson}, \& {Lewin}}]{2008bassa}
{Bassa}, C.~G., {Pooley}, D., {Verbunt}, F., {et~al.} 2008, \aap, 488, 921

\bibitem[{{Becker}(2009)}]{2009becker}
{Becker}, W. 2009, in Astrophysics and Space Science Library, Vol. 357,
  Astrophysics and Space Science Library, ed. W.~{Becker}, 91--140

\bibitem[{{Blandford} \& {Teukolsky}(1976)}]{1976blandford}
{Blandford}, R. \& {Teukolsky}, S.~A. 1976, \apj, 205, 580

\bibitem[{{Burderi} {et~al.}(2002){Burderi}, {D'Antona}, \&
  {Burgay}}]{2002burderi}
{Burderi}, L., {D'Antona}, F., \& {Burgay}, M. 2002, \apj, 574, 325

\bibitem[{{Cheng} {et~al.}(2006){Cheng}, {Taam}, \& {Wang}}]{cheng2006}
{Cheng}, K.~S., {Taam}, R.~E., \& {Wang}, W. 2006, \apj, 641, 427

\bibitem[{{D'Amico} {et~al.}(2001{\natexlab{a}}){D'Amico}, {Lyne},
  {Manchester}, {Possenti}, \& {Camilo}}]{damico2001}
{D'Amico}, N., {Lyne}, A.~G., {Manchester}, R.~N., {Possenti}, A., \& {Camilo},
  F. 2001{\natexlab{a}}, \apjl, 548, L171

\bibitem[{{D'Amico} {et~al.}(2001{\natexlab{b}}){D'Amico}, {Possenti},
  {Manchester}, {Sarkissian}, {Lyne}, \& {Camilo}}]{damico2001b}
{D'Amico}, N., {Possenti}, A., {Manchester}, R.~N., {et~al.}
  2001{\natexlab{b}}, \apjl, 561, L89

\bibitem[{{Edmonds} {et~al.}(2003{\natexlab{a}}){Edmonds}, {Gilliland},
  {Heinke}, \& {Grindlay}}]{2003edmonds1}
{Edmonds}, P.~D., {Gilliland}, R.~L., {Heinke}, C.~O., \& {Grindlay}, J.~E.
  2003{\natexlab{a}}, \apj, 596, 1177

\bibitem[{{Edmonds} {et~al.}(2003{\natexlab{b}}){Edmonds}, {Gilliland},
  {Heinke}, \& {Grindlay}}]{2003edmonds2}
{Edmonds}, P.~D., {Gilliland}, R.~L., {Heinke}, C.~O., \& {Grindlay}, J.~E.
  2003{\natexlab{b}}, \apj, 596, 1197

\bibitem[{{Ferraro} {et~al.}(2001){Ferraro}, {Possenti}, {D'Amico}, \&
  {Sabbi}}]{2001ferraro}
{Ferraro}, F.~R., {Possenti}, A., {D'Amico}, N., \& {Sabbi}, E. 2001, \apjl,
  561, L93

\bibitem[{{Grindlay} {et~al.}(2002){Grindlay}, {Camilo}, {Heinke}, {Edmonds},
  {Cohn}, \& {Lugger}}]{2002grindlay}
{Grindlay}, J.~E., {Camilo}, F., {Heinke}, C.~O., {et~al.} 2002, \apj, 581, 470

\bibitem[{{Grindlay} {et~al.}(2001){Grindlay}, {Heinke}, {Edmonds}, {Murray},
  \& {Cool}}]{2001grindlay}
{Grindlay}, J.~E., {Heinke}, C.~O., {Edmonds}, P.~D., {Murray}, S.~S., \&
  {Cool}, A.~M. 2001, \apjl, 563, L53

\bibitem[{{Orosz} \& {van Kerkwijk}(2003)}]{2003orosz}
{Orosz}, J.~A. \& {van Kerkwijk}, M.~H. 2003, \aap, 397, 237

\bibitem[{{Pallavicini} {et~al.}(1981){Pallavicini}, {Golub}, {Rosner},
  {Vaiana}, {Ayres}, \& {Linsky}}]{1981pallavicini}
{Pallavicini}, R., {Golub}, L., {Rosner}, R., {et~al.} 1981, \apj, 248, 279

\bibitem[{{Possenti} {et~al.}(2005){Possenti}, {D'Amico}, {Corongiu},
  {Manchester}, {Sarkissian}, {Camilo}, \& {Lyne}}]{2005possenti}
{Possenti}, A., {D'Amico}, N., {Corongiu}, A., {et~al.} 2005, in Astronomical
  Society of the Pacific Conference Series, Vol. 328, Binary Radio Pulsars, ed.
  {F.~A.~Rasio \& I.~H.~Stairs}, 189

\bibitem[{{Sabbi} {et~al.}(2003){Sabbi}, {Gratton}, {Bragaglia}, {Ferraro},
  {Possenti}, {Camilo}, \& {D'Amico}}]{2003sabbi}
{Sabbi}, E., {Gratton}, R.~G., {Bragaglia}, A., {et~al.} 2003, \aap, 412, 829

\bibitem[{{Verbunt} {et~al.}(2008){Verbunt}, {Pooley}, \&
  {Bassa}}]{2008verbunt}
{Verbunt}, F., {Pooley}, D., \& {Bassa}, C. 2008, in IAU Symposium, Vol. 246,
  IAU Symposium, 301--310

\end{thebibliography}

\end{document}